\documentstyle[11pt,newpasp,twoside,psfig]{article}
\def\edcomment#1{\iffalse\marginpar{\raggedright\sl#1\/}\else\relax\fi}
\reversemarginpar

\newcommand{\kms}          {\mbox{${\rm km~s^{-1}}$}}
\newcommand{\cc}           {\mbox{${\rm cm^{-3}}$}}
\newcommand{\e}            {\mbox{$^{-1}$}}
\newcommand{\ee}           {\mbox{$^{-2}$}}

\def\oversim#1#2{\lower0.5ex\vbox{\baselineskip=0pt\lineskip=0.2ex
     \ialign{$\mathsurround=0pt #1\hfil##\hfil$\crcr#2\crcr\sim\crcr}}}
\def\simgt{\mathrel{\mathpalette\oversim>}}

\def\nh2{\mbox{$n_{\rm H_2}$}}
\def\Nh2{\mbox{$N_{{\rm H}_2}$}}
\def\hcop{\mbox{HCO$^+$}}
\def\h13cop{\mbox{H$^{13}$CO$^+$}}

\begin{document}
\title{The formation of stars in clusters}
\author{Jonathan Williams}
\affil{Astronomy Department, University of Florida, Gainesville, FL 32611}

\begin{abstract}
Observations of the dust and gas around embedded stellar clusters
reveal some of the processes involved in their formation and evolution.
Large scale mass infall with rates
$\dot M\simeq 4\times 10^{-4}~M_\odot$~yr\e\
is found to be disrupted on small scales by protostellar outflows. 
Observations of the size and velocity dispersion of clusters suggest
that protostellar migration from their birthplace begins at very early
times and is a potentially useful evolutionary indicator.
\end{abstract}

\section{Introduction}

Massive stars form in the densest regions of the most massive
molecular clouds and they do so in groups. To study massive
star formation, therefore, requires that we learn about the
processes by which stars form in clusters.

Much of what we have learnt about star formation, however, has come
from studying nearby, isolated, low mass stars (e.g. Evans 1999).
Protostellar evolution passes through four stages, Class 0 to III,
that are defined on the basis of their observed spectral energy
distribution and is physically related to the mass of the molecular
envelope around the star (Andr\'{e}, Ward-Thompson, \& Barsony 2000).
As a protostar grows through accretion while simultaneously
dispersing the surrounding gas through outflows,
the column density of the envelope around it
decreases and the emergent spectral energy distribution shifts
to shorter wavelengths from far-infrared (Class 0) to visible (Class III).
The earliest stages are the shortest lived, with statistically
based estimates of the lifetime of the Class 0 stage
$\sim 5\times 10^4$~yr, and Class I $\sim 10^5$~yr
compared to Class II and III lifetimes $> 10^6$~yr
(Mundy, Looney, \& Welch 2000).

Just as the availability of infrared arrays facilitated the study
of clusters of Class II sources in the 1990s (e.g., Lada 1992),
so the arrival of millimeter and sub-millimeter wavelength bolometer
arrays has made possible large scale studies of clusters at the
earlier Class 0--I stages of protostellar evolution
(e.g. Johnstone, this volume)
We are now in a position to address questions about cluster
formation and evolution with a level of detail that approaches
the case of isolated star formation.

There are important differences between the formation of stars in
clusters and in isolation. For instance, cluster forming clouds
may initiate collapse through the dissipation of turbulent support
(Nakano 1998) rather than the quasi-static gravitational contraction
of a magnetically supported core (Shu, Adams, \& Lizano 1987)
and may form stars more efficiently (Lada 1992).
Massive stars in particular, drastically alter their environment
(e.g., Walborn, this volume)
and may induce further star formation through cloud compression
(Elmegreen \& Lada 1977). On the other hand, they may also be the
principal agents of cloud destruction (Williams \& McKee 1997).

To learn about the processes by which clouds gather the material
to form clusters and the effect of newly formed stars back on the
cloud requires observations of the intra-cluster gas and dust.
Here I discuss radio observations of two highly embedded,
Class 0 and I protostellar groups and present a potential new
method for following cluster evolution at these early stages.

\section{Millimeter wavelength observations of NGC2264}

NGC2264 lies in the northern region of Monoceros, 760~pc distant,
and contains about 300 near-infrared and 30 IRAS sources.
The former are mostly Class II protostars and the latter Class I.
Using the Heinrich Hertz Telescope on Mt. Graham, AZ we observed
dust continuum emission at $870~\mu$m and $J=3-2$ lines of
\hcop\ and \h13cop\ toward three protostellar groups in NGC2264
to search for redder, younger sources and to follow the motions
of the gas around them. Details of the observing procedure and
data analysis can be found in Williams \& Garland (2001).

The location of the observations is shown in Figure~1.
IRS1 is a bright, compact group centered on an early B star
situated just north of the Cone nebula.
IRS2 and IRS3 are less luminous IRAS dust condensations lying
$\sim 6'$ and $11'$, respectively, to the north.

\vskip 0.1in
\centerline{\psfig{figure=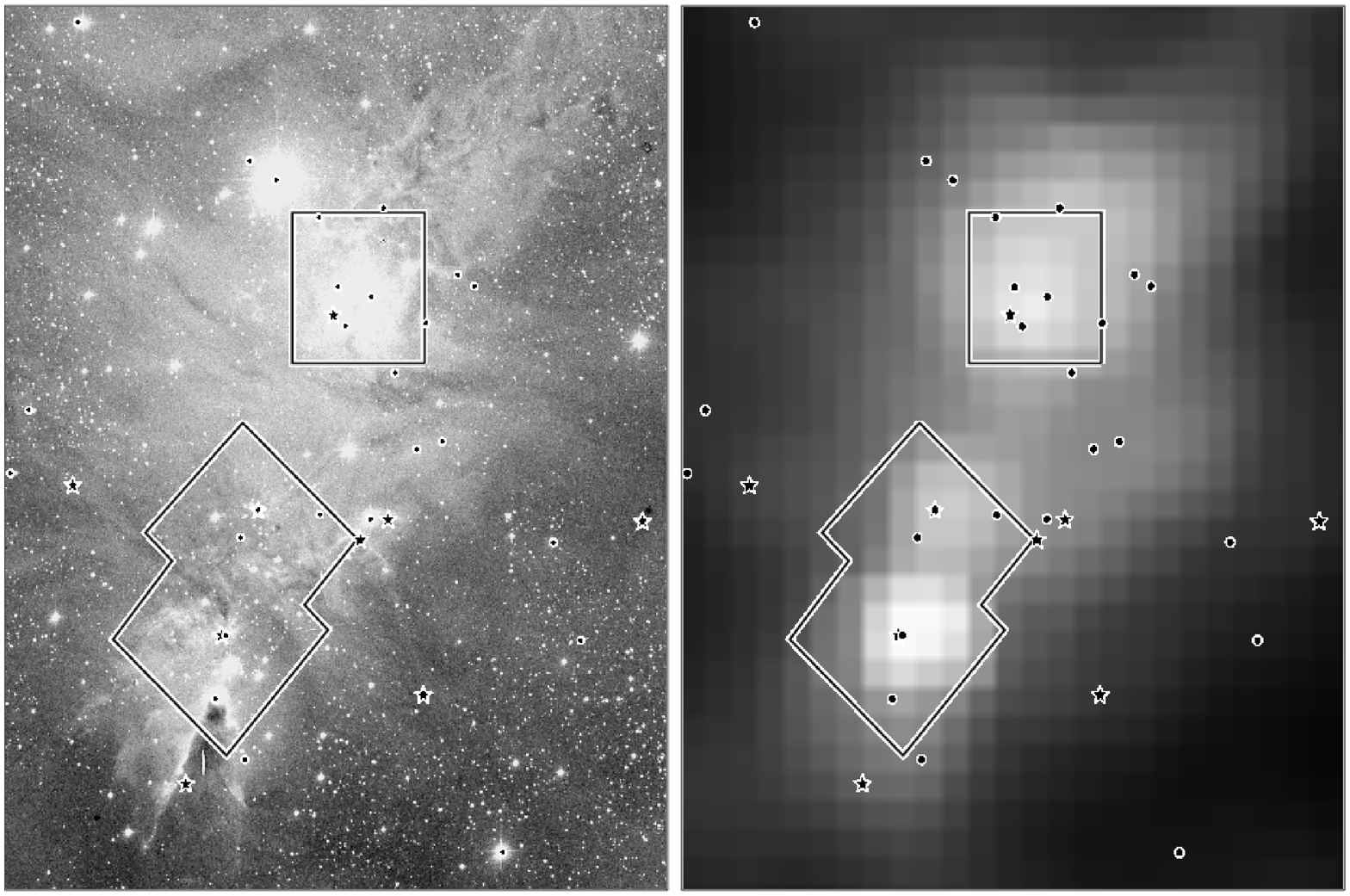,height=3.4in,angle=90,silent=1}}
\vskip 0.1in
\begin{small}
\noindent{\bf Figure 1:}
Digital sky survey and IRAS $100~\mu$m images of the NGC2264 cluster
forming region. Color selected IRAS and MSX point sources (stars and
circles respectively) and the boundaries of the HHT $870~\mu$m bolometer
maps are overlaid on each image.
\end{small}
\bigskip

\subsection{Structure}

Thermal dust emission at $870~\mu$m toward IRS1 and IRS2 is shown in
Figure~2. IRS1 is the bright, compact source to the south. Its elongated
shape shows signs of substructure; Ward-Thompson et al. (2000)
resolve five distinct condensations with higher
resolution IRAM and JCMT observations. IRS2 lies to the north
and is more fragmented with several low intensity peaks, suggestive of
a slightly more evolved cluster in which the individual protostars are
beginning to shed their circumstellar envelopes and migrate away from
their birthplace. IRS3 was not detected in a short integration and is
only included here in speculative comments toward the end of the paper.

\vskip 0.1in
\centerline{\psfig{figure=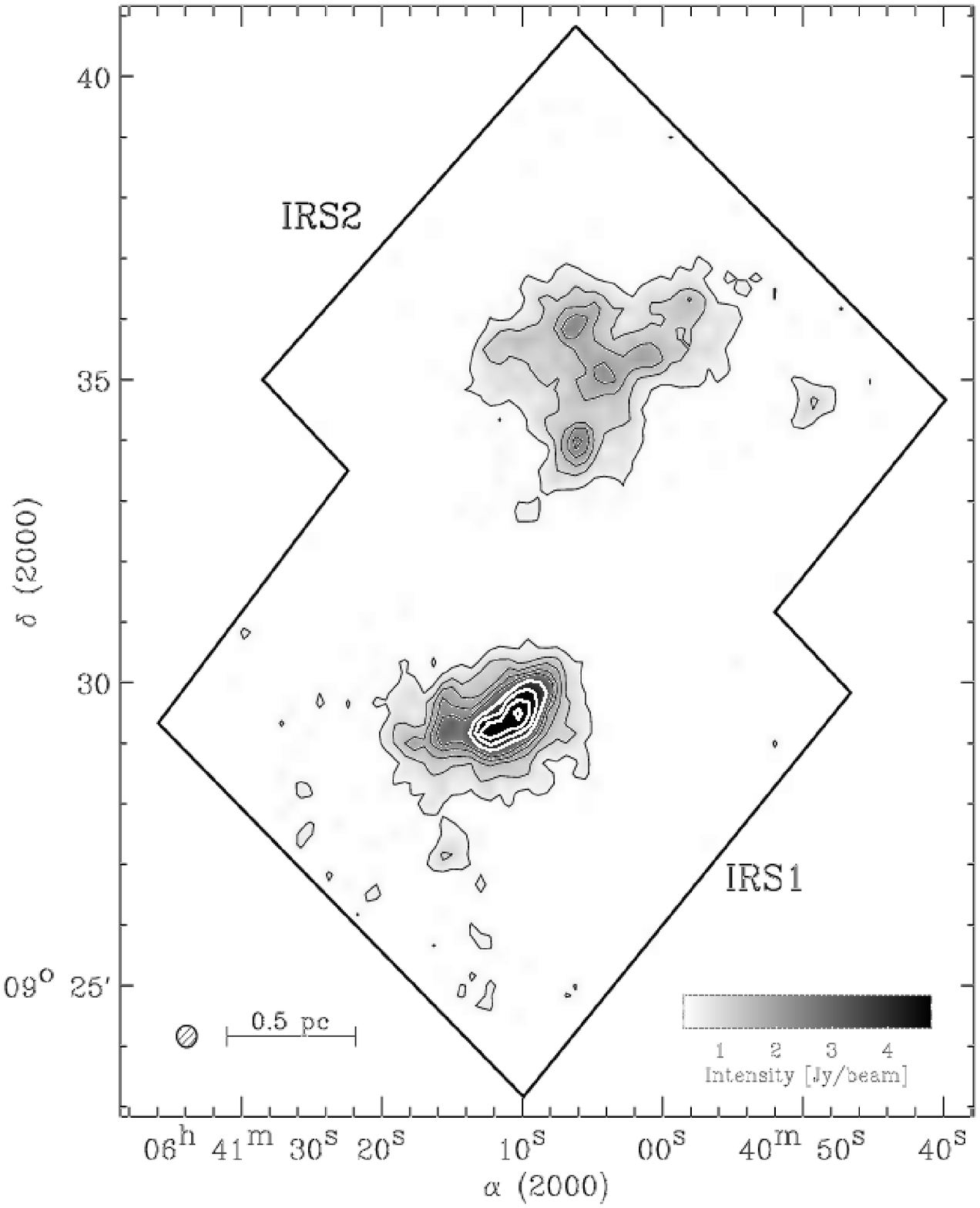,height=4.0in,angle=0,silent=1}}
\vskip 0.1in
\begin{small}
\noindent{\bf Figure 2:}
$870~\mu$m continuum emission toward IRS1 and IRS2.
Black contours are at 0.5, 1.0, 1.5, 2.0, 2.5 Jy~beam\e,
and white contours are at 3.5, 4.5, 5.5, 6.5 Jy~beam\e.
The boundary of the map is marked by the heavy solid line and the
$21''$ beam size is indicated in the lower left corner.
\end{small}
\bigskip

The integrated fluxes, for flux densities greater than 0.5~Jy~beam\e,
of the two clusters are similar, 94~Jy for IRS1 and 84~Jy for IRS2
indicating that they have similar dust masses and make a good
comparative study. This implies total (gas + dust) masses of
$790~M_\odot$ (IRS1) and $710~M_\odot$ (IRS2) for the envelopes
around each protostellar group, for a mass opacity
$\kappa=0.009$~cm$^2$~g\e\ and dust temperature $T_d=17$~K,
The projected area of each group,
0.45~pc$^2$ for IRS1 and 0.62~pc$^2$ for IRS2,
implies average column and volume densities,
$\langle\Nh2\rangle=8\times 10^{22}$~cm\ee,
$\langle\nh2\rangle=5\times 10^4$~\cc\ for IRS1 and
$\langle\Nh2\rangle=5\times 10^{22}$~cm\ee,
$\langle\nh2\rangle=3\times 10^4$~\cc\ for IRS2.

The high column densities toward the protostars imply
$A_{\rm V}\simgt 50$ and suggest they are indeed young
(Class 0--I) objects.
The average density of the envelopes traced by the continuum
observations is similar to the critical density of \hcop.
Further, integrated line maps show a similar morphology to the
dust (Williams \& Garland 2001) and we therefore use the line
data to investigate the dynamics of the gas around the clusters
and the protostars within. Each group is discussed separately:
the $29''\simeq 0.1$~pc resolution of the line data is too coarse
to investigate the dynamics around the individual protostars
in IRS1 and we focus on global properties for this group;
in the larger IRS2 group, individual objects are resolved
and the gas motions around each one are analyzed.

\subsection{Dynamics}

Both the optically thin \h13cop\ and optically thick \hcop\ $J=3-2$
lines were observed. The former allows us to analyze the systemic
velocity, linewidth, and abundance in the clumps around each cluster
and the latter provides information on relative motions,
i.e. infall and outflow, in the gas.
Because of the limited space, I can only discuss the infall motions
here and refer the reader to Williams \& Garland (2001) for the
\h13cop\ velocity and abundance analysis.

The \hcop\ spectra generally show two peaks and a central dip at
the mean velocity of \h13cop. Such self-absorbed profiles can be
used to diagnose relative motions in the gas (e.g., Walker et al. 1986).
Average spectra are shown toward IRS1 in Figure~3. The averages
were computed over different regions of the clump corresponding
to different $870~\mu$m flux densities. This is similar to radial
averaging but takes into account the non-circular shape of the clump.
At low flux densities, corresponding to the outer regions,
the average spectra are relatively weak and the optical depths
are low. Nevertheless, the \hcop\ spectra are clearly asymmetric
and show the red-shifted self-absorption characteristic of collapse.
At higher flux densities, toward the clump center, the
line intensities and optical depths increase and the central
self-absorption dip becomes more prominent. However, the
spectra also become more symmetric with increasing flux density
suggesting a decrease in the net inflow.

Determining the infall speed profile requires detailed modeling.
We ran the 1-d radiative transfer code, RATRAN, of
Hogerheijde \& van der Tak (2000) and used the continuum and \h13cop\
observations to fix the size, temperature, density profile, abundance,
mean velocity and dispersion. Only the relative velocity profile and
[$^{12}$C]/[$^{13}$C] isotope ratio remain as variables in the
the modeled output \hcop\ line profiles.

Model spectra are shown as dotted lines in Figure~3 and reproduces
the qualitative features in the \hcop\ spectra reasonably well given
the tight constraints from the dust and \h13cop\ observations.
The model consists of a spherical shell at a radius $r_{\rm in}=0.32$~pc,
with mass $400~M_\odot$, infalling at speed $v_{\rm in}=0.3$~\kms\
onto an expanding inner layer moving outwards at
$v_{\rm exp}=1.0$~\kms. The abundance of \hcop\ was constant
and equal to $1.5\times 10^{-9}$ relative to H$_2$,
similar to values in other OB star forming regions (Bergin 1997).
The inferred mass infall rate onto the cluster is,
$$\dot M_{\rm in} = M_{\rm in}v_{\rm in}/r_{\rm in}
                  = 4\times 10^{-4}~M_\odot {\rm yr}^{-1}.$$
This is one to two orders of magnitude greater than infall rates
onto solar and intermediate mass protostars Zhou (1995)
as expected for a global infall onto a group of such objects
and is similar to the value expected for a gravitational flow,
$\dot M=\sigma^3/G$, where $\sigma=1.0$~\kms\ is the average velocity
dispersion of the core and $G$ is the gravitational constant (Shu 1977).

\vskip -0.3in
\centerline{\psfig{figure=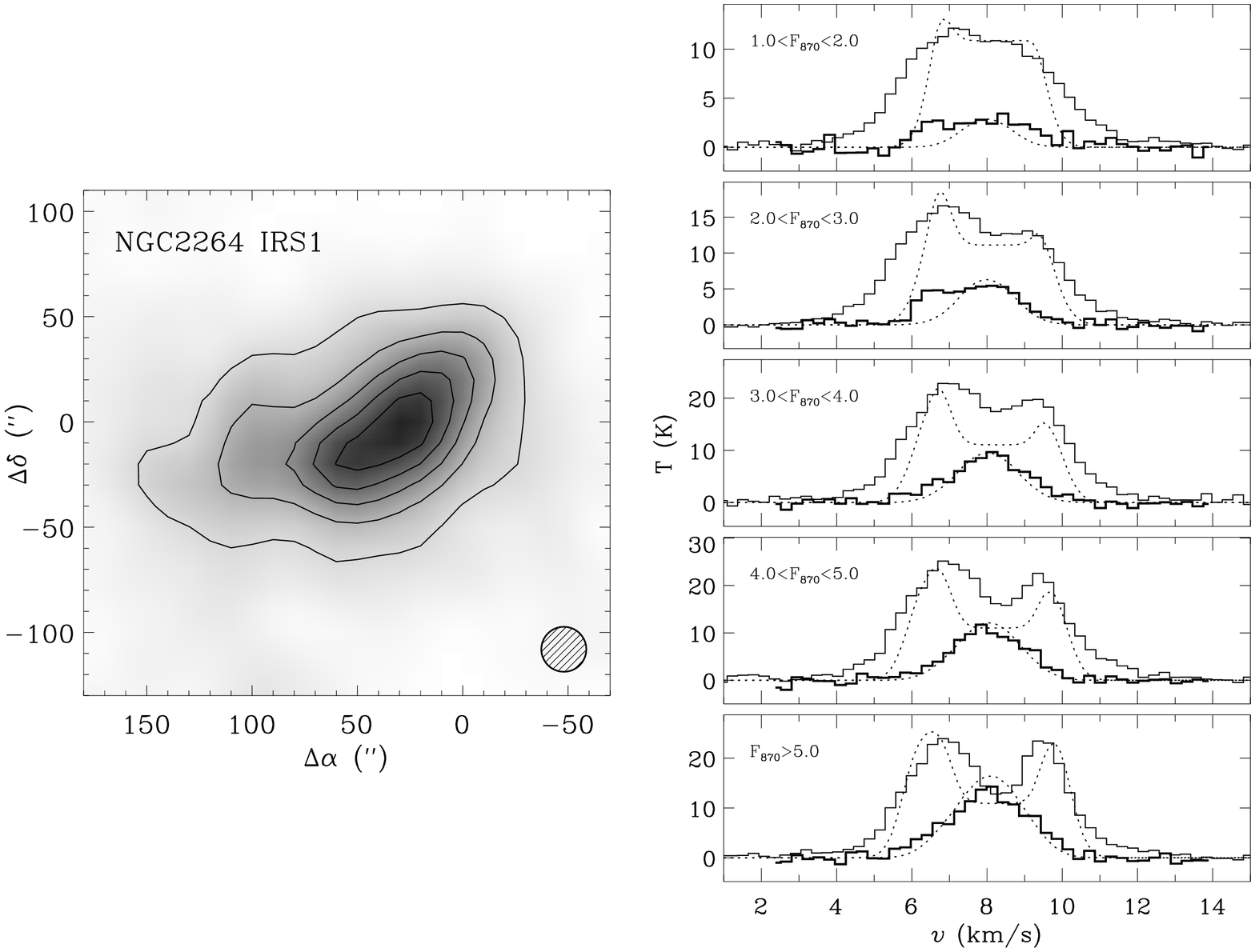,height=4.6in,angle=90,silent=1}}
\vskip -0.15in
\begin{small}
\noindent{\bf Figure 3:}
Global properties in IRS1. The left panel shows the continuum emission
smoothed to the $29''$ resolution of the line data. Contours are at
1,2,3,4,5~Jy~beam\e, and spectra averaged within each contour interval
are shown on the right hand side (\hcop\ in light, \h13cop\ in dark
and multiplied by 3 for clarity).
At low flux densities/large radii, the \hcop\ optical depth is low
and the self-absorption effect is small but a clear asymmetry can be
seen indicating collapse. At high flux densities/small radii, the
self-absorption dip is strong but the spectra are more symmetric
indicating a smaller net infall rate. Model fits, as described in
the text, are shown as dotted lines.
\end{small}
\bigskip

The expanding inner region is likely due to the protostellar outflows
in the region, one of which appears to be directed along our line of
sight (Schreyer et al. 1997). Shu et al. (1987) hypothesize
that (low mass) stars self-determine their final mass through the removal
of a collapsing envelope by their outflows.
By comparing clusters in different evolutionary states
and examining the gas flows around them, it will be possible to see
if such a process is relevant in a clustered, massive star forming,
environment, and therefore assess its role in the origin of the stellar
IMF (Adams \& Fatuzzo 1996).

The average \hcop\ spectra toward the IRS2 cluster also shows red-shifted
self-absorption and model fits imply a similar infall speed and mass infall
rate (Williams \& Garland 2001). However, the cluster is more widely
dispersed and its more irregular appearance precludes a similar radial
average study as in IRS1.
On the other hand, the greater separation of the protostars
allows an investigation of the flows around each one individually.
Spectra toward each of six identified cores are shown in Figure~4.

\vskip -0.8in
\centerline{\psfig{figure=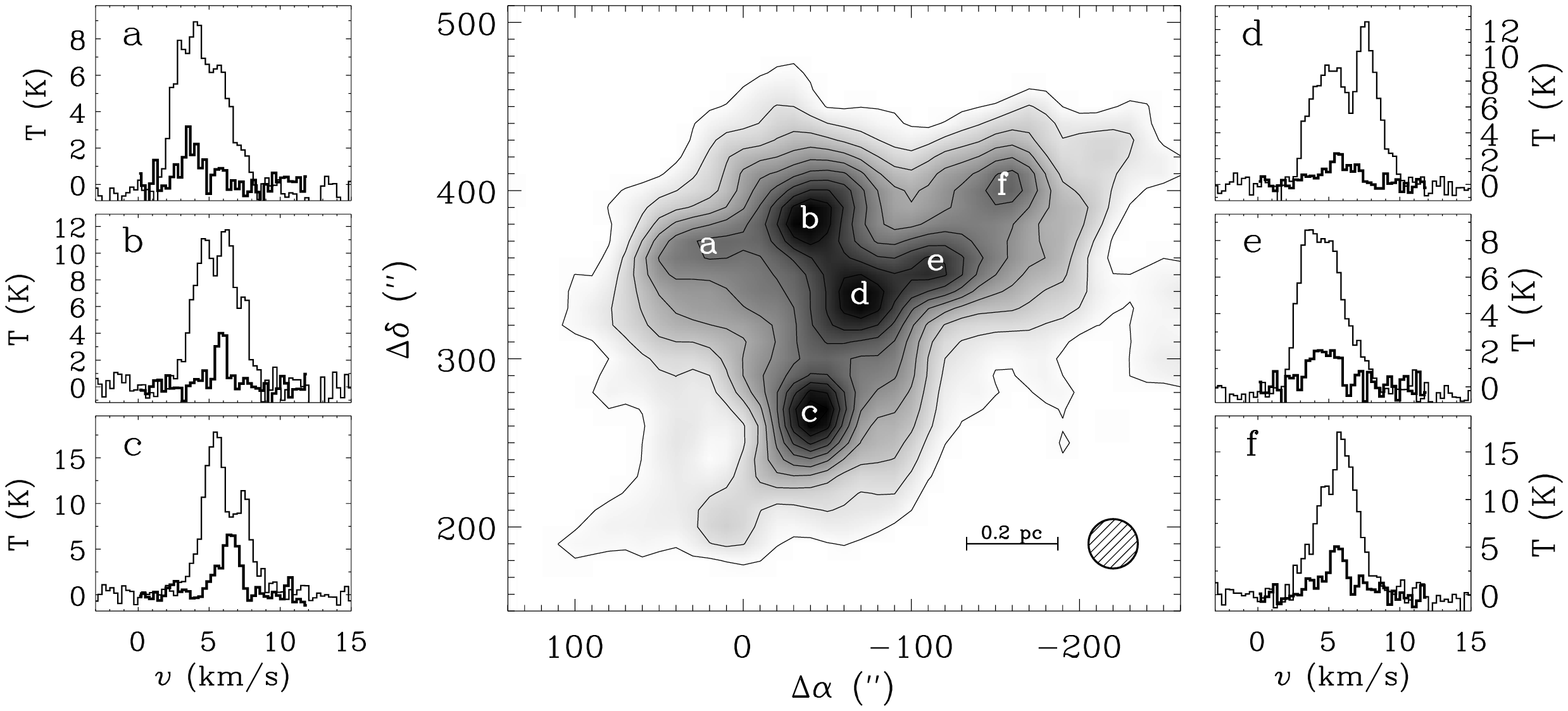,height=4.6in,angle=90,silent=1}}
\vskip -0.9in
\begin{small}
\noindent{\bf Figure 4:}
Spectra toward individual protostars in IRS2. The central panel shows
the continuum emission smoothed to the spectral line resolution of $29''$.
Contours begin at and are in steps of 0.2 Jy~beam\e.
Six distinct peaks of emission are labeled a-f and spectra
(light: \hcop; dark: \h13cop, multiplied by 3 for clarity)
toward each one are shown in the panels on either side.
\end{small}
\bigskip

Core c shows the clearest infall signature with a deep, red-shifted,
absorption dip. A model fit using the RATRAN code
implies an infall speed $v_{\rm in}=0.3$~\kms, and mass
infall rate $\dot M_{\rm in} = 2\times 10^{-5}~M_\odot {\rm yr}^{-1}$.
The speed is the same as the collapse onto the entire cluster
but $\dot M$ is a factor of 3 less than the total
mass infall rate divided by the number of protostars:
apparently not all the material falling onto the cluster makes it
all the way to the scale of protostars. This is similar to the
case in IRS1 but here we witness the effect on a star-by-star
basis as in core d which has a reversed (outflow) spectrum.

The six cores have a dispersion $\sigma_{\rm c-c}=0.90$~\kms\
and extend over an area with an equivalent circular radius $r=0.44$~pc.
Together, these imply a virial mass,
$M_{\rm virial}=3r\sigma_{\rm c-c}^2/G=250~M_\odot$.
This is less than the total interstellar mass estimated
from the continuum observations but is likely much greater
than the final stellar mass of the system (Lada et al. 1993).
Thus, although the system is currently bound, as the surrounding
gas is dispersed through, e.g., stellar outflows, it is likely to
become unbound (Hills 1980).

An unbound cluster expands and its age may be estimated from its
size and expansion rate. This effect is seen in optically visible
clusters in Orion over a range of ages $\sim 3-12$~Myr (Blaauw 1991).
The morphology of the two groups in Figure~2 hint that this process
may begin even in the protostellar stage: the relatively low column
densities toward IRS2 suggest that it is more evolved than IRS1 and
offer a natural explanation for its greater size.
Working backwards from its size and velocity dispersion implies
a kinematic age, $t=r/\sigma_{\rm c-c}=5\times 10^5$~yr,
characteristic of late Class I sources.
Given the continued infall onto the cluster and the large
envelope mass, it is not clear if the cluster is actually expanding
and whether this simple age estimate is a useful measure
at these early stages in a cluster's formation.
However, it offers the following simple, testable, picture of
cluster evolution.

\section{A possible scenario for cluster evolution}

Do clusters evolve from a compact state such as IRS1 to a more
extended state such as IRS2 as the circumcluster (and circumstellar)
envelope is dispersed? Additional observations of other embedded
stellar groups in NGC2264 and other regions will show if there is
a consistent pattern between cluster age (as judged by the SEDs
of the cluster and protostars within) and morphology. In this regard,
it is noteworthy that the widely separated group of MSX sources
in IRS3 (see Figure~1) was not detected at $870~\mu$m to the same
noise level (0.1~Jy~beam\e) as IRS1 and 2.
Thus we might imagine the cartoon model of cluster
evolution in Figure~5.

\vskip -0.25in
\centerline{\psfig{figure=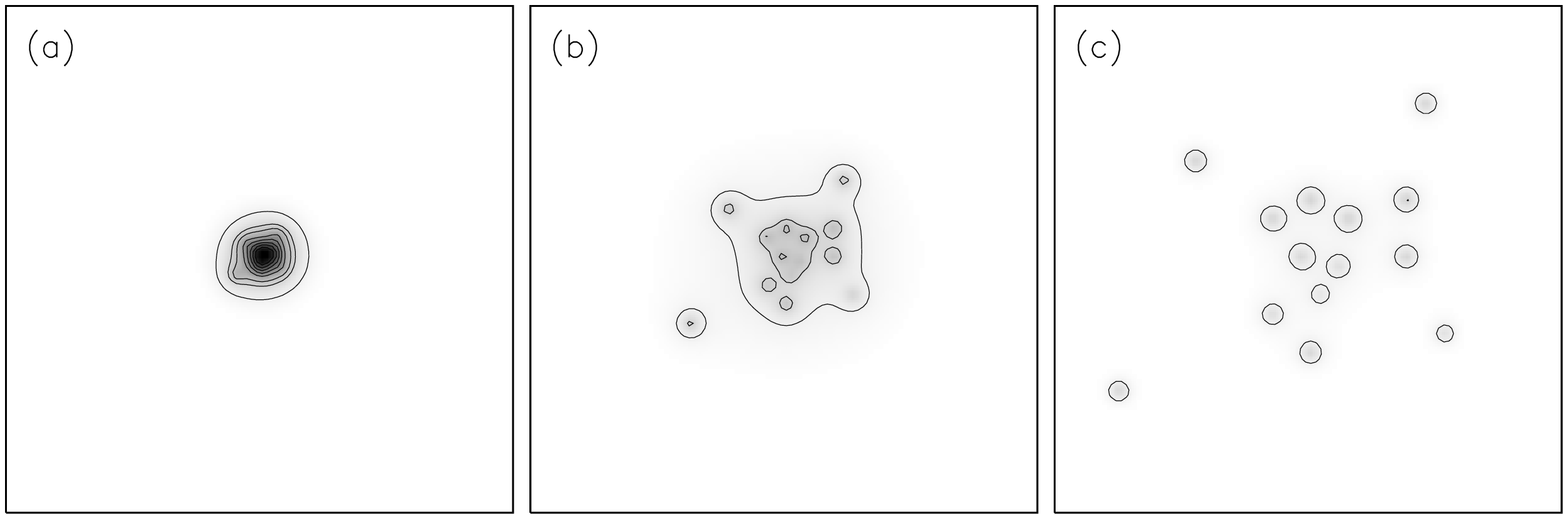,height=4.6in,angle=90,silent=1}}
\vskip -2.3in
\begin{small}
\noindent{\bf Figure 5:}
Schematic model of cluster evolution. This simple toy model consists
of a group of stars expanding at a constant velocity embedded in
individual protostellar envelopes and surrounded by a circumcluster
envelope that spreads out linearly with time.
The left panel represents a stellar group at early (Class 0) times:
protostars are closely spaced and deeply embedded within a common,
compact envelope. As the cluster evolves, the envelope begins to
disperse and the (Class I) protostars migrate away from each other
although there is still substantial circumcluster emission.
At later stages, in the rightmost panel, the circumcluster gas
has mostly dispersed, and the stars are widely separated, surrounded
only by relatively low mass (Class II) protostellar envelopes.
\end{small}
\bigskip

Observations of the early, bright, compact phase of cluster evolution
is well suited to interferometry. In later stages, large scale mapping
with sub-millimeter bolometer arrays are necessary to image the more
extended groups and targeted radio observations to detect remnants of
low mass cores around more evolved sources (Saito et al. 2001)
can provide essential kinematic information to determine the
gravitational stability of a group and estimate its age.

Studies of young clusters are essential for understanding the way in
which most, and particularly massive, stars form. The above cartoon
model best applies to the accretion model of massive star formation
(Maeder; McKee, this volume) but the agglomeration model (Bonnell,
this volume) requires an earlier stage, lasting for $\sim 10^5$~yr,
consisting of a large group of low mass protostars that first contracts
before creating massive stars. Searches for such protoclusters provide
a test of the competing theories.

\bigskip

The Heinrich Hertz Telescope is operated by the Submillimeter Telescope
Observatory on behalf of Steward Observatory and the Max-Planck-Institut
fuer Radioastronomie. I thank Michiel Hogerheijde and Floris van der Tak
for their advice in the use of the RATRAN program.

\end{document}